\documentclass[10pt,conference]{IEEEtran}

\usepackage[utf8]{inputenc}
\usepackage{amsmath,amsfonts,amssymb}
\usepackage{graphicx}
\usepackage{float}
\usepackage{booktabs}
\usepackage{multirow}
\usepackage{array}
\usepackage{xcolor}
\usepackage{subcaption}
\usepackage{algorithm}
\usepackage{algorithmic}
\usepackage{cite}
\usepackage{hyperref}

\hypersetup{
    colorlinks=true,
    linkcolor=black,
    citecolor=black,
    urlcolor=blue,
    pdfauthor={Anonymous},
    pdftitle={},
    pdfproducer={},
    pdfcreator={}
}

\title{Quantum Optimization for Access Point Selection Under Budget Constraint}

\author{

    \IEEEauthorblockN{Mohamed Khalil Brik}
    \IEEEauthorblockA{
        Computer Science and Engineering\\
        American University in Cairo\\
        Cairo, Egypt\\
        mohamedkhalil.brik@aucegypt.edu
    }
    \and
    \IEEEauthorblockN{Ahmed Shokry}
    \IEEEauthorblockA{
        Computer Science and Engineering\\
        Pennsylvania State University\\
        PA, USA\\
        ahmed.shokry@psu.edu 
    }
    \and
    \IEEEauthorblockN{Moustafa Youssef}
    \IEEEauthorblockA{
        Computer Science and Engineering\\
        American University in Cairo\\
        Cairo, Egypt\\
        moustafa-youssef@aucegypt.edu
    }

}

\begin{document}

\maketitle

\begin{abstract}
\boldmath
Optimal Access Point (AP) selection is crucial for accurate indoor localization, yet it is constrained by budget, creating a trade-off between localization accuracy and deployment cost. Classical approaches to AP selection are often computationally expensive, hindering their application in large-scale 3D indoor environments.

In this paper, we introduce a quantum APs selection algorithm under a budget constraint. The proposed algorithm leverages quantum annealing to identify the most effective subset of APs allowed within a given budget. We formulate the APs selection problem as a quadratic unconstrained binary optimization (QUBO) problem, making it suitable for quantum annealing solvers. The proposed technique can drastically reduce infrastructure requirements with a negligible impact on performance.

We implement the proposed quantum algorithm and deploy it in a realistic 3D testbed. Our results show that the proposed approach can reduce the number of required APs by \textbf{96.1\%} while maintaining a comparable 3D localization accuracy. Furthermore, the proposed quantum approach outperforms classical AP selection algorithms in both accuracy and computational speed. Specifically, our technique achieves a time of \textbf{0.20 seconds}, representing a speedup of 61$\times$ over its classical counterpart, while reducing the mean localization error by \textbf{10\%} compared to the classical counterpart. 
For floor localization, the quantum approach achieves \textbf{73\%} floor accuracy, outperforming both the classical AP selection (\textbf{58.6\%}) and even using the complete set of APs (\textbf{70.4\%}). This highlights the promise of the proposed quantum APs selection algorithm for large-scale
3D localization.

\end{abstract}


\section{Introduction}
\label{sec:introduction}

Indoor positioning systems have become increasingly critical for applications ranging from navigation assistance to asset tracking and emergency response~\cite{zafari2019survey}. While Global Positioning System (GPS) technology provides excellent outdoor localization accuracy, indoor environments present unique challenges including signal attenuation, multipath interference, and the absence of satellite connectivity~\cite{liu2007survey}. Consequently, WiFi-based fingerprinting has emerged as the dominant approach for indoor localization, leveraging the ubiquity of wireless access points (APs) in modern buildings~\cite{bahl2000radar, youssef2005horus, he2016wifi, deak2012survey}. Leveraging Radio Frequency (RF) technology, WiFi-based fingerprinting systems function through a two-stage procedure. The initial offline phase involves constructing a radio map, or \textit{RF fingerprint} by systematically collecting Received Signal Strength (RSS) data from transmitters such as WiFi access points, cell towers, and Bluetooth beacons at known locations across the site. Subsequently, in the online phase, the RSS readings from a device at an unknown location are gathered and matched against the pre-established fingerprint database to determine the most probable location based on signal similarity.

Traditional indoor localization research has primarily focused on discrete floor classification or two-dimensional (2D) positioning within known floor levels~\cite{xia2017indoor, elbakly2018truestory}. However, the demands of practical applications have evolved, now requiring continuous three-dimensional (3D) localization that can simultaneously determine a device's latitude, longitude, and floor with high accuracy~\cite{guo20173d, luo2017indoor, elbakly2018truestory}. This shift to 3D reinforces the underlying optimization challenge, which involves selecting an optimal subset of exactly \(k\) access points from a pool of \(n\) candidates to maximize accuracy.
This challenge is further compounded by the dense deployment of APs in modern buildings. While this density offers a theoretical opportunity for improved precision, it paradoxically creates significant practical hurdles. The high number of APs introduces signal redundancy, increases computational complexity, and inflates hardware and maintenance costs~\cite{he2015efficient, kaemarungsi2004distribution}. Consequently, current approaches that utilize all available APs without systematic optimization result in over-provisioned systems that suffer from high costs for only marginal performance gains~\cite{li2015access}.

Recently, there have been several studies addressing the Access Point (AP) selection problem to enhance indoor localization systems. Classical techniques have been extensively explored, with methods such as using a modified Jaccard index~\cite{zhou2018jaccard}, employing PCA and autoencoders for feature extraction \cite{jiang2021fingerprint}, and applying mutual information for dynamic AP pruning \cite{mi_online_ap_selection}. These classical approaches, while effective at establishing a trade-off between accuracy and cost, are often computationally expensive for large-scale deployments. On the other hand, emerging quantum techniques have been proposed to overcome these computational hurdles by formulating the problem for quantum annealers \cite{shokry2024quantum}. However, a key limitation of existing quantum formulations is that they do not incorporate a strict budget constraint in their model. This omission leaves them exposed to selecting a number of APs that could exceed practical available hardware limitations, despite their potential for speed.

In this paper, we introduce a quantum AP selection algorithm that operates under a strict budget constraint, addressing the key challenges in resource-constrained 3D indoor localization. Our approach formulates the AP selection problem as a Quadratic Unconstrained Binary Optimization (QUBO) problem with cardinality constraints. The proposed quantum algorithm utilizes binary variables to denote AP selection and formulates an objective function with a dual purpose: to prioritize APs that have a strong effect on location information (i.e., maximize importance) and to deprioritize those with high correlation to others already selected (i.e., minimize redundancy), all while strictly adhering to the selection of exactly \(k\) APs. This formulation makes it suitable for solving via quantum annealing~\cite{mcgeoch2014adiabatic}. A core component of our method is the comparison of different importance metrics for weighting APs within the QUBO framework. These metrics, based on analyzing the signal strength of each AP across different locations, include signal entropy, signal variance, signal average, and signal max. This comparison allows us to identify the most effective strategy for quantifying an AP's importance.

We implement the proposed quantum algorithm and deploy it in a realistic 3D testbed. Our results demonstrate that this technique can drastically reduce infrastructure requirements, cutting the number of required APs by 96.1\% with a negligible impact on performance. The proposed quantum approach outperforms classical AP selection algorithms in both accuracy and computational speed, achieving a solution in 0.20 seconds, a speedup of $61\times$ over its classical counterpart, while reducing the mean localization error by \textbf{10\%} compared to the classical counterpart. For floor localization, the quantum approach achieves \textbf{73\%} floor accuracy, outperforming both the classical AP selection  (\textbf{58.6\%}) and even using the complete set of APs (\textbf{70.4\%}). This performance highlights the promise of our quantum optimization framework for enabling large-scale, cost-effective 3D localization systems.

The rest of the paper is organized as follows: Section~\ref{sec:background} gives a background on  quantum annealing and the fingerprinting localization problem. Section~\ref{sec:method} provides the details of our quantum APs selection algorithm. We present the details of the implementation and evaluation results in Section~\ref{sec:evaluation}. Finally, Section \ref{sec:conclude} concludes the paper.

\section{Background}
\label{sec:background}
In this section, we present the foundational concepts necessary to understand our proposed approach. We first provide a concise overview of quantum annealing. Then, we introduce the fundamentals of fingerprinting-based 3D indoor localization, which constitutes the application domain of our work.

\subsection{Quantum Annealing}

Quantum annealing is a specialized paradigm within quantum computing~\cite{mcgeoch2014adiabatic, kadowaki1998quantum}, designed for solving complex optimization problems that involve identifying the best solution among an exponentially large number of possibilities. It follows the principles of adiabatic quantum computation~\cite{farhi2000quantum, albash2018adiabatic}, where a quantum system is initialized in the ground state of a simple Hamiltonian \(H_0\) and gradually evolved into the ground state of a more complex Hamiltonian \(H_f\)~\cite{aharonov2008adiabatic}. This adiabatic evolution enables the system to reach the optimal solution encoded in \(H_f\). The initial Hamiltonian \(H_0\) represents a simple, easily prepared state, while \(H_f\) encodes the target optimization problem~\cite{johnson2011quantum}.

This process exploits quantum phenomena such as superposition and tunneling~\cite{muthukrishnan2016tunneling} to explore the solution space more effectively than classical algorithms, potentially escaping local minima. The system's evolution during annealing is governed by the time-dependent Schr\"{o}dinger equation~\cite{griffiths2018introduction}:
\begin{equation}
i\hbar\frac{\partial}{\partial t}|\psi(t)\rangle = H(t)|\psi(t)\rangle
\end{equation}
where \(i\) is the imaginary unit, \(\hbar\) the reduced Planck constant, \(H(t)\) the time-dependent Hamiltonian operator, and \(|\psi(t)\rangle\) the system's quantum state at time \(t\). The Hamiltonian is typically expressed as a linear combination of \(H_0\) and \(H_f\), controlled by a time-dependent parameter \(s(t)\) that varies from 0 to 1~\cite{santoro2002theory}:
\begin{equation}
H(t) = (1 - s(t))H_0 + s(t)H_f
\end{equation}

This construction is central not only to quantum mechanics but also to combinatorial optimization~\cite{lucas2014ising}. Many real-world optimization problems can be expressed as Quadratic Unconstrained Binary Optimization (QUBO) formulations~\cite{kochenberger2014unconstrained, glover2022tutorial}, defined as:
\begin{equation}
\text{Minimize } f(\mathbf{x}) = \sum_{i,j} Q_{ij}x_ix_j + \sum_i P_ix_i
\end{equation}
where \(\mathbf{x}\) is a binary vector, \(Q_{ij}\) are coefficients of the quadratic terms, and \(P_i\) are coefficients of the linear terms~\cite{boros2002pseudo}.

QUBO problems can be reformulated as an Ising Hamiltonian, a model that represents magnetic interactions among atomic spins~\cite{brush1967history, barahona1982computational}. In this mapping, binary variables correspond to spin states, and the QUBO objective function maps onto the Ising energy function~\cite{choi2008minor}. This equivalence allows quantum annealers such as those developed by D-Wave to solve QUBO problems natively~\cite{boixo2014evidence, dwave2020advantage}. Quantum annealing has been applied across diverse domains including finance~\cite{orus2019quantum}, logistics~\cite{neukart2017traffic}, machine learning~\cite{nath2021review}, and network optimization~\cite{venturelli2015quantum}.

\subsection{Fingerprinting in Indoor Localization}
\begin{figure}[!t]
    \centering
    \includegraphics[width=\linewidth]{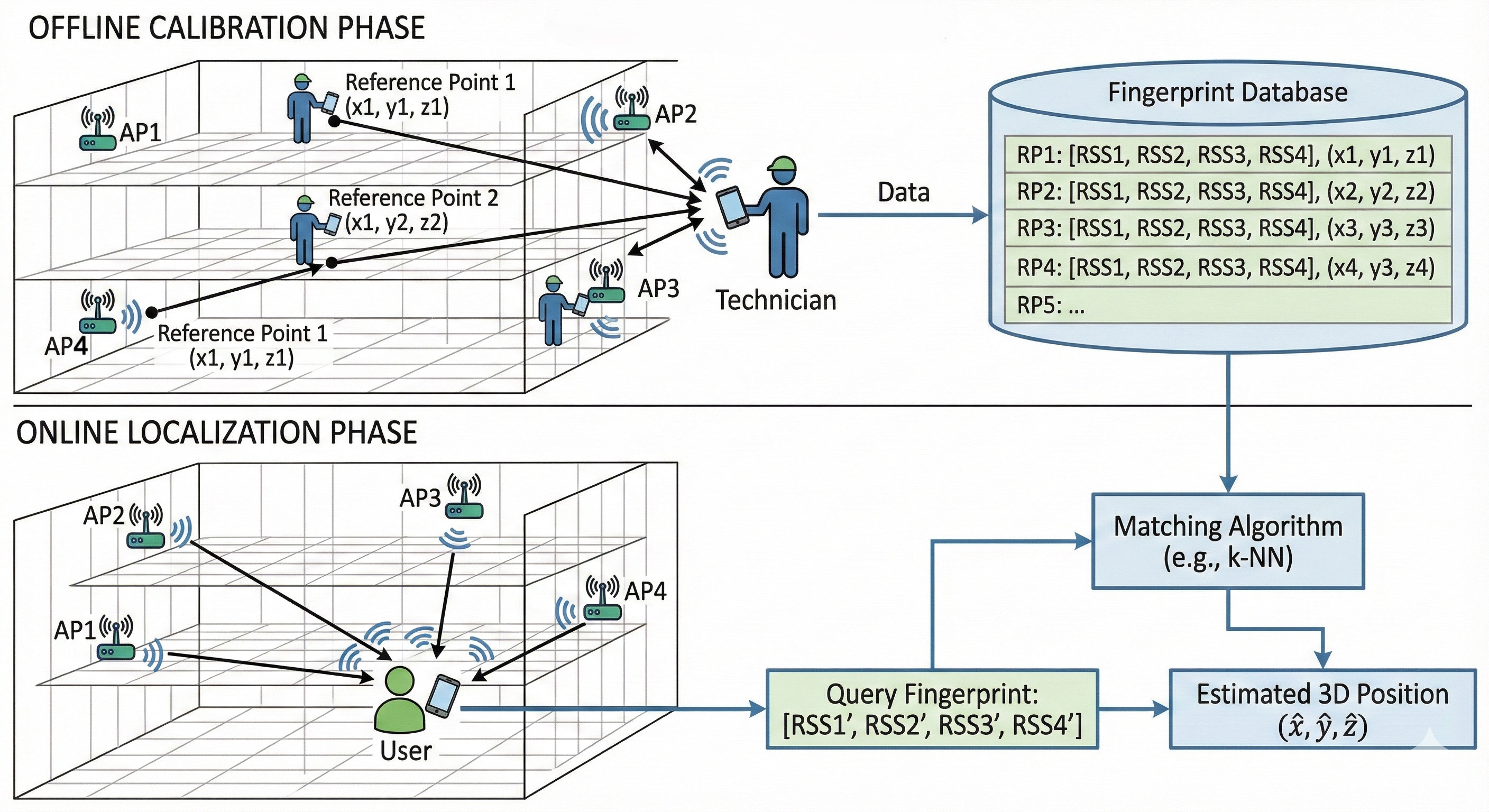}
    \caption{Example of WiFi fingerprinting location tracking process with four APs.}
    \label{fig:fingerprinting}
\end{figure}

Three-dimensional (3D) indoor localization aims to determine a user's precise position, including horizontal coordinates and vertical height, within complex indoor environments~\cite{guo20173d, shang2022overview}. Among various approaches, fingerprinting has emerged as a leading technique due to its robustness against multipath effects and signal attenuation~\cite{he2016wifi, safwat2023fingerprinting}. Fingerprinting-based 3D localization typically operates in two phases as shown in Figure~\ref{fig:fingerprinting}. This figure shows a representative 3D indoor environment with multiple access points (AP1, AP2, AP3, AP4) deployed across different floors. 

In the \textit{offline calibration phase}, a technician systematically collects RSS measurements from all visible APs at known reference points throughout the 3D space, recording each location's coordinates (x, y, z) along with the corresponding RSS vector. These fingerprints are stored in a database where each entry contains the RSS readings from all APs and the associated 3D position. In the \textit{online localization phase}, a user's mobile device measures RSS values from nearby APs to form a query fingerprint, which is then matched against the pre-built database using a similarity metric (e.g., k-nearest neighbors). The position corresponding to the best-matching fingerprint is returned as the estimated 3D location ~\cite{he2016wifi, safwat2023fingerprinting}.

\section{Methodology}
\label{sec:method}
This section details our proposed AP selection algorithm. We begin by formulating the budget-constrained AP selection problem as a Quadratic Unconstrained Binary Optimization (QUBO) problem. The formulation is then broken down into its core components: a term to maximize the selection of influential APs, a term to minimize redundancy among the selected APs, and a constraint term to enforce the strict AP budget. The key notations used throughout this section are summarized in Table~\ref{tab:notations}.

\begin{table}[!t]
\centering
\caption{Table of Notations}
\label{tab:notations}
\begin{tabular}{|c|l|}
\hline
\textbf{Symbol} & \textbf{Description} \\ \hline
\( Q(\mathbf{x}, \alpha, \eta) \) & The QUBO objective function. \\ \hline
\( \mathbf{x} \) & A binary AP selection vector of size \(n\). \\ \hline
\( x_i \in \mathbf{x}\) & \(x_i = 1\) if AP \(i\) is selected, \(0\) otherwise. \\ \hline
\( \alpha \) & The balancing parameter (\(0 \le \alpha \le 1\)). \\ \hline
\( n \) & The total number of candidate APs. \\ \hline
\( m \) & The total number of samples in the fingerprint. \\ \hline
\( k \) & The desired number of APs to select. \\ \hline
\( \eta \) & The penalty coefficient for the constraint. \\ \hline
\( I_i \) & The calculated importance score for AP \(i\). \\ \hline
\( R_{ij} \) & The redundancy score between APs \(i\) and \(j\). \\ \hline
\( r_i, r_j \) & Vectors of RSS measurements for APs \(i\) and \(j\)\\ \hline
\( r_{i,l} \) & The \(l\)-th RSS measurement for AP \(i\). \\ \hline
\( \bar{r}_i \) & The mean of the RSS vector for AP \(i\). \\ \hline
\end{tabular}
\end{table}

\subsection{QUBO Objective}

Given a 3D fingerprint $\mathcal{D} = \{ (r^i, y^i) \}_{i=1}^{m}$, with $m$ samples and $n$ access points. For each sample $i \in \{1, \dots, m\}$, the observed measurement from $n$ APs consists of an RSS vector $r^i \in \mathbb{R}^n$ and a 3D fingerprint location label $y^i \in \mathbb{R}^3$, where each label $y^i$ combines latitude, longitude, and floor height. 
The goal is to select exactly $k$ APs (AP budget constraint), encoded in a binary selection vector $\mathbf{x} = (x_1, ..., x_n)\in\{0, 1\}^n$ subject to
\begin{equation}
\sum_{j=1}^{n} x_j = k
\label{eq:selection_constraint}
\end{equation}
such that localization accuracy using the reduced fingerprint vector $r^i_S \in \mathbb{R}^k$ remains maximized. To this end, we define the following Quadratic Unconstrained Binary Optimization (QUBO) objective:

\begin{multline}
Q(\mathbf{x}, \alpha) = \\
-\alpha \sum_{i=1}^{n} I_i x_i + (1-\alpha) \sum_{i=1}^{n} \sum_{j>i} R_{ij} x_i x_j + \eta \left(\sum_{i=1}^{n} x_i - k\right)^2
\end{multline}

The QUBO objective function \( Q(\mathbf{x}, \alpha) \) is defined over a binary selection vector \( \mathbf{x} \), where each element \( x_i \) indicates whether access point \( i \) is selected (\( x_i = 1 \)) or not (\( x_i = 0 \)). The term \( I_i \) represents the weighted importance score of access point \( i \), quantifying its individual contribution to localization accuracy; this forms the importance component of the objective, weighted by the parameter \( \alpha \). The redundancy between access points \( i \) and \( j \) is captured by \( R_{ij} \), which measures the correlation between their signals; this constitutes the redundancy part of the objective, scaled by \( 1-\alpha \) to balance against importance. The parameter \( \alpha \) thus controls the trade-off between maximizing the total importance of selected access points and minimizing their mutual redundancy. To enforce the selection of exactly \( k \) access points, the penalty term \( \eta \left(\sum_{i=1}^{n} x_i - k\right)^2 \) is included, where \( \eta \) is an adaptive weight that ensures the constraint is satisfied.

The optimal subset, $\mathbf{x}^*$, is then obtained as:
\begin{equation}
\mathbf{x}^* =
\arg\min_{\substack{\mathbf{x} \in \{0, 1\}^n \\[2pt] \sum_{j=1}^{n} x_j = k}}
Q(\mathbf{x}, \alpha, \eta).
\label{eq:qubo_minimization}
\end{equation}

This formulation enables integration with modern quantum and classical QUBO solvers, supporting scalable optimization under practical deployment constraints.

\subsection{Importance}
The importance of an AP is determined by analyzing the variation of its Received Signal Strength (RSS) across different locations in the radio map. Intuitively, an AP whose RSS remains constant throughout the environment provides no discriminative information and is therefore unimportant for localization. Conversely, an AP whose RSS varies significantly between locations is highly valuable for distinguishing one position from another.

\subsubsection{\textbf{Entropy-Based Importance}} This metric measures the Shannon entropy of each AP's RSS distribution, quantifying the variability in signal strength readings. For AP \(i\) with RSS probability distribution \(p(r_{i,l})\) for all samples $l\in m$ (at all locations), the importance is defined as:
\begin{equation}
I_i^{\text{ENT}} = -\sum_{l\in m} p(r_{i,l}) \log_2 p(r_{i,l})
\end{equation}
Higher entropy indicates more diverse signal patterns, potentially useful for distinguishing locations. However, this metric does not explicitly consider the relationship with spatial coordinates.

\subsubsection{\textbf{Variance-Based Importance}} The variance metric captures the spread of RSS measurements for each AP across all fingerprint locations:
\begin{equation}
I_i^{\text{VAR}} = \frac{1}{m-1}\sum_{l=1}^{m} (r_{i,l} - \bar{r}_i)^2
\end{equation}
where \(\bar{r}_i\) is the mean RSS for AP \(i\). APs with high variance exhibit location-dependent signal characteristics, while low-variance APs provide little discriminative information.

\subsubsection{\textbf{Average-Based Importance}} This metric uses the mean received signal strength as a proxy for AP $i$ importance:
\begin{equation}
I_i^{\text{AVG}} = \bar{r_i} = \frac{1}{m}\sum_{l=1}^{m} r_{i,l}
\end{equation}
This approach assumes that APs with stronger average signals are more reliable and contribute more to localization accuracy, though it ignores spatial variation patterns.

\subsubsection{\textbf{Maximum-Based Importance}} This metric identifies APs based on their peak signal strength:
\begin{equation}
I_i^{\text{MAX}} = \max_{l=1,\ldots,m} r_{i,l}
\end{equation}
It assumes that APs with the highest maximum signal strength have better coverage and signal quality, potentially leading to more accurate localization.

\subsection{Redundancy}

To quantify the pairwise redundancy between access points (APs), we compute a correlation matrix using the Pearson correlation coefficient applied to the received signal strength RSS data. Specifically, we consider only those APs with non-zero importance scores to ensure relevance. For each pair of relevant APs \(i\) and \(j\), the absolute value of the Pearson correlation coefficient between their RSS vectors is calculated, producing a symmetric redundancy matrix \(R\) where each element \(R_{ij}\) represents the degree of redundancy between the corresponding APs. High values of \(R_{ij}\) indicate strong correlation and thus potential redundancy, which the selection algorithm aims to minimize to avoid overlapping or dependent information contributions.

\begin{align}
R_{ij} &= \left| \text{Corr}(r_i, r_j) \right| \\
&= \left|
\frac{\sum_{l=1}^{m} (r_{i,l} - \bar{r}_i)(r_{j,l} - \bar{r}_j)}{\sqrt{\sum_{l=1}^{m}(r_{i,l} - \bar{r}_i)^2 \sum_{l=1}^{m}(r_{j,l} - \bar{r}_j)^2}}
\right|
\end{align}


\subsection{Budget Constraint}
The parameter \( \alpha \) governs the trade-off between two objectives: maximizing the total importance of the selected access points and minimizing their mutual redundancy. The formulation ensures that both objectives are balanced for any given value of \( \alpha \). However, tuning \( \alpha \) alone does not guarantee that exactly \( k \) APs will be selected. To enforce this budget constraint precisely, we introduce a penalty term \( \eta \left( \sum_{i=1}^{n} x_i - k \right)^2 \), where \( \eta \) is a sufficiently large weight that ensures the solution satisfies the requirement of selecting exactly \( k \) APs.

\section{Evaluation}
\label{sec:evaluation}

In this section, we evaluate our proposed quantum algorithm and implement it on a quantum machine simulator~\cite{openjij2022}. We start by describing our real testbed. Then, we show the effect of the different system parameters. Finally, we compare the proposed quantum algorithm with its classical counterpart. Table~\ref{tab:default_params} shows our system parameters and their default values.
 
\begin{table}[!t]
\centering
\caption{System parameters and their default values}
\label{tab:default_params}
\begin{tabular}{ll}
\toprule
\textbf{Parameter} & \textbf{Default Value} \\
\midrule
AP Budget ($k$) & 20 \\
Importance-Redundancy Trade-off ($\alpha$) & 0.8 \\
Penalty Coefficient($\eta$) & 2.0 \\
Importance Metric & Entropy \\
Number of Reads & 1,000 \\
Number of Sweeps & 1,000 \\
Inverse Temperature ($\beta$) & 10.0 \\
Transverse Field ($\gamma$) & 1.0 \\
\bottomrule
\end{tabular}
\end{table}

\subsection{Experiment Setup}

The evaluation utilized a publicly available dataset covering a five-story building with a total area of approximately $110\textrm{m}^{2}$~\cite{torres2014ujiindoorloc}. The dataset incorporates signals from 520 WiFi access points comprising both the building's own infrastructure and overheard APs from adjacent buildings. 
A fingerprinting database of $m=21,048$ samples was used for the experiments. Spatial coordinates (latitude, longitude) are normalized to [0,1], while floor indices remain discrete integers. We set the floor height to 3.0 meters for 3D Euclidean distance calculations. For the floor classification task, a Random Forest Classifier was employed. All classical benchmarks were executed on a workstation equipped with a 12th Gen Intel Core i7-12700H processor running at 2.30 GHz with 16 GB RAM, operating on Windows 11 (64-bit). 

For solving the QUBO formulation, we employ OpenJij~\cite{openjij2022} which is an open-source framework for quantum annealing simulation that implements quantum Monte Carlo methods. The quantum annealing parameters are set to 1,000 reads (independent optimization runs) and 1,000 sweeps (Monte Carlo steps per run), ensuring sufficient exploration of the solution space. In the next subsections, we perform comprehensive hyperparameter optimization across QUBO parameters ($k$, $\alpha$, $\eta$) and annealing-specific parameters (\texttt{num\_sweeps}, inverse temperature $t$) to identify optimal configurations. 
Performance is evaluated using 3D localization error, floor classification accuracy, and time-to-solution (TTS) metrics, with all results representing mean and standard deviation over multiple independent trials. 
As a comparison baseline, we utilize Simulated Annealing (SA) sampler~\cite{dwave2022}, which implements classical simulated annealing without quantum tunneling effects.

\subsection{Importance}



\begin{figure}[!t]
    \centering
    \includegraphics[width=0.99\linewidth]{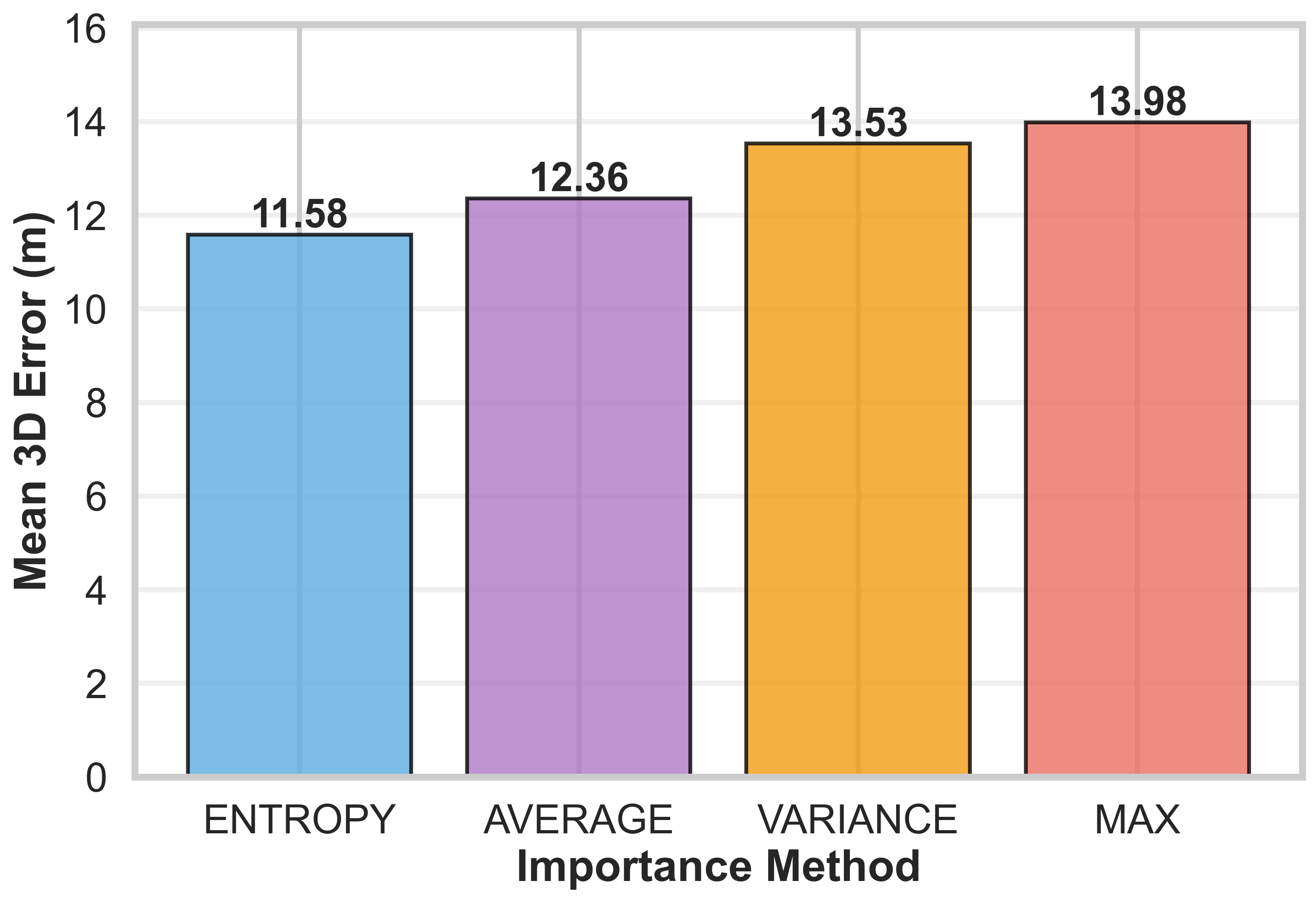}
    \caption{Mean 3D error by selection method.}
    \label{fig:method_vs_error}
\end{figure}

\begin{figure}[!t]
    \centering
    \includegraphics[width=0.99\linewidth]{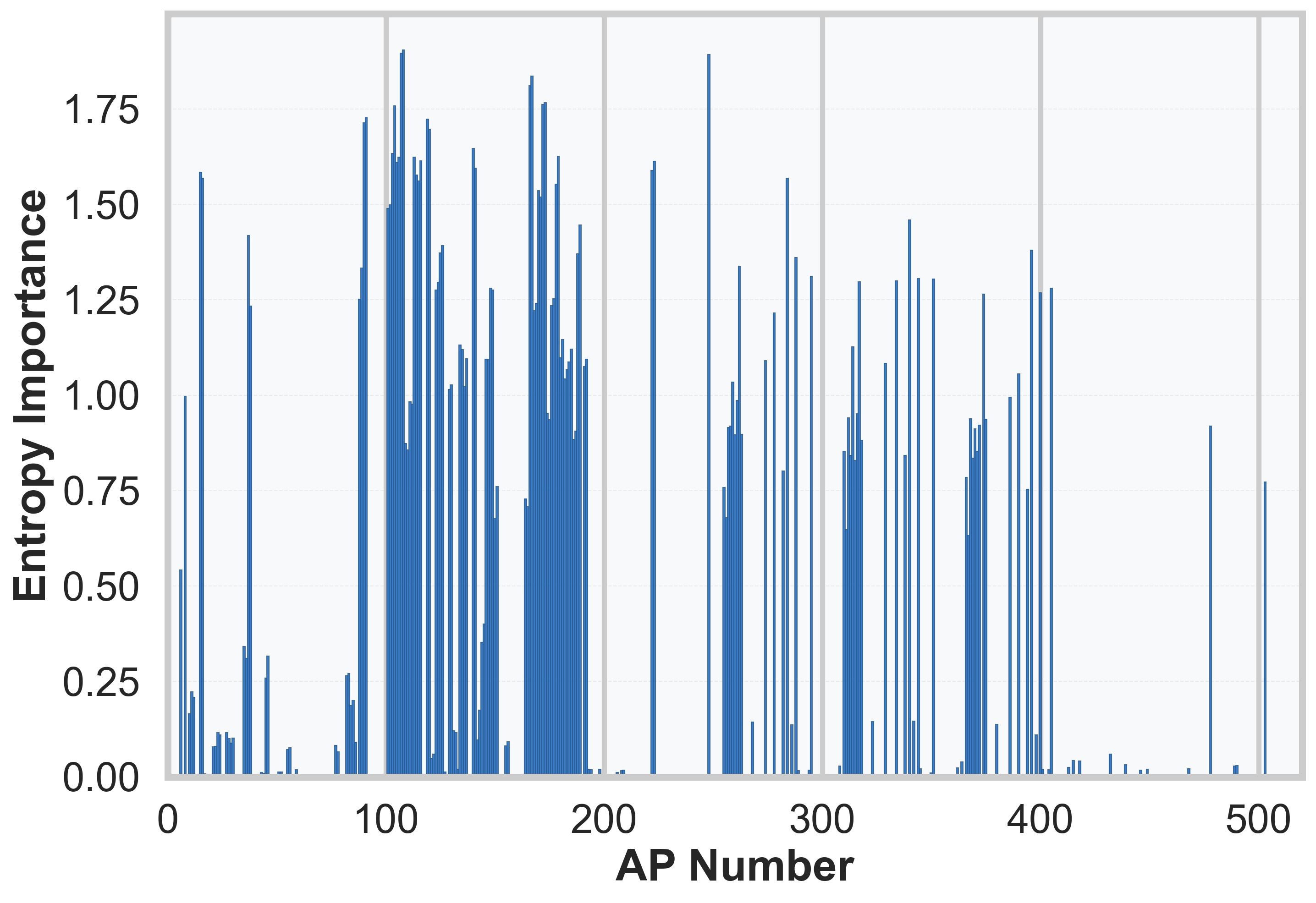}
    \caption{Entropy Importance.}
    \label{fig:entropy_importanc}
\end{figure}

Figure~\ref{fig:method_vs_error} compares the performance of the proposed four importance metrics for AP selection. The figure shows that entropy-based importance achieves the best mean 3D localization error of 11.58m, demonstrating that signal variability measured through Shannon entropy most effectively identifies informative APs. This result indicates that for fingerprinting-based localization, consistent signal patterns across diverse locations are more valuable than peak signal strength or high variability alone. Figure~\ref{fig:entropy_importanc} further shows the Entropy importance score computed for every access point in the environment. The distribution reveals that a substantial number of APs contribute very little to the localization model. Pruning these low-importance APs can significantly reduce the system's computational complexity while having a negligible impact on localization accuracy.

\subsection{Redundancy}

\begin{figure}[!t]
    \centering
    \includegraphics[width=0.99\linewidth]{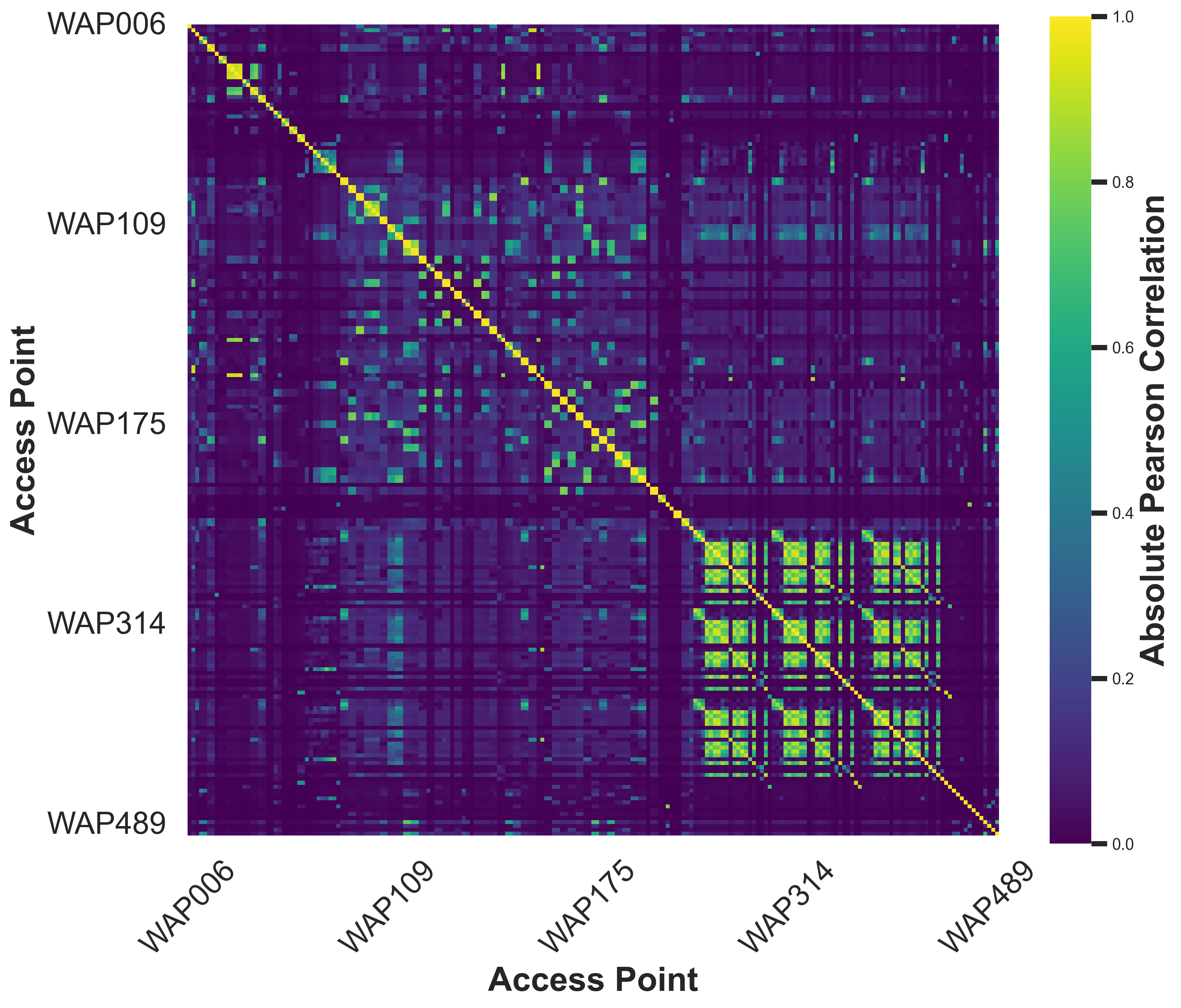}
    \caption{Pairwise correlation matrix of AP RSS patterns.}
    \label{fig:redundancy}
\end{figure}

Figure~\ref{fig:redundancy} visualizes the correlation matrix for the access points, revealing significant redundancy among numerous AP pairs (indicated by the yellow regions). This correlation introduces extra computational overhead without yielding a corresponding improvement in localization accuracy.

\subsection{Effect of QUBO Parameters}

\subsubsection{Effect of Balancing Parameter \texorpdfstring{$\alpha$}{alpha}}

\begin{figure}[!t]
    \centering
    \includegraphics[width=0.99\linewidth]{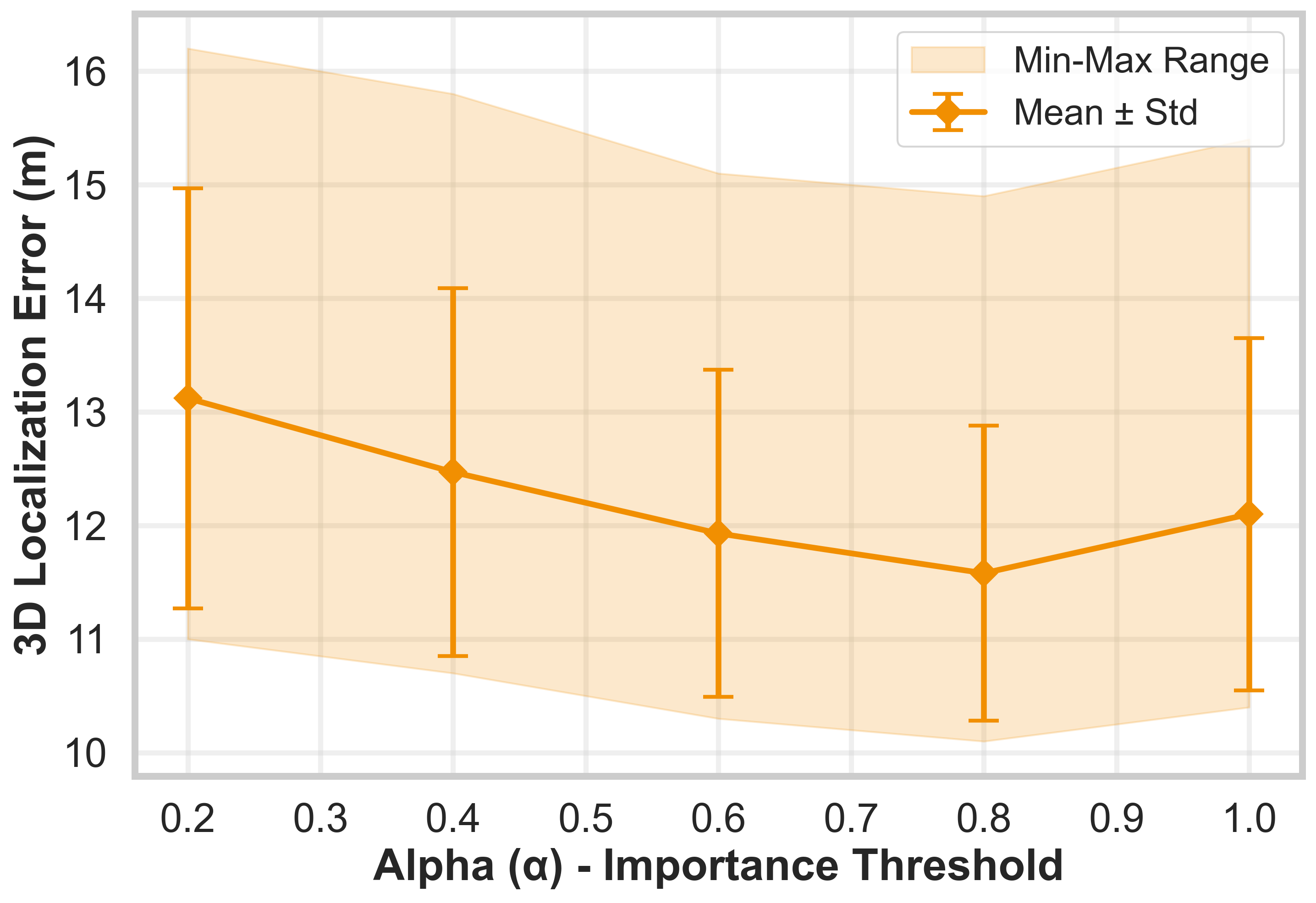}
    \caption{Effect of \texorpdfstring{$\alpha$}{alpha} parameter on mean 3D localization error.}
    \label{fig:alpha_vs_error}
\end{figure}

Figure~\ref{fig:alpha_vs_error} illustrates the impact of the $\alpha$ parameter on 3D localization accuracy, where $\alpha$ controls the trade-off between importance maximization and redundancy minimization in the QUBO objective. The figure suggests that heavily weighting importance (e.g., at $\alpha=0.80$) produces consistent average results, which demonstrates that the QUBO formulation effectively balances both the importance and redundancy.

\subsubsection{Effect of Penalty Parameter \texorpdfstring{$\eta$}{eta}}
\begin{figure}[!t]
    \centering
    \includegraphics[width=0.99\linewidth]{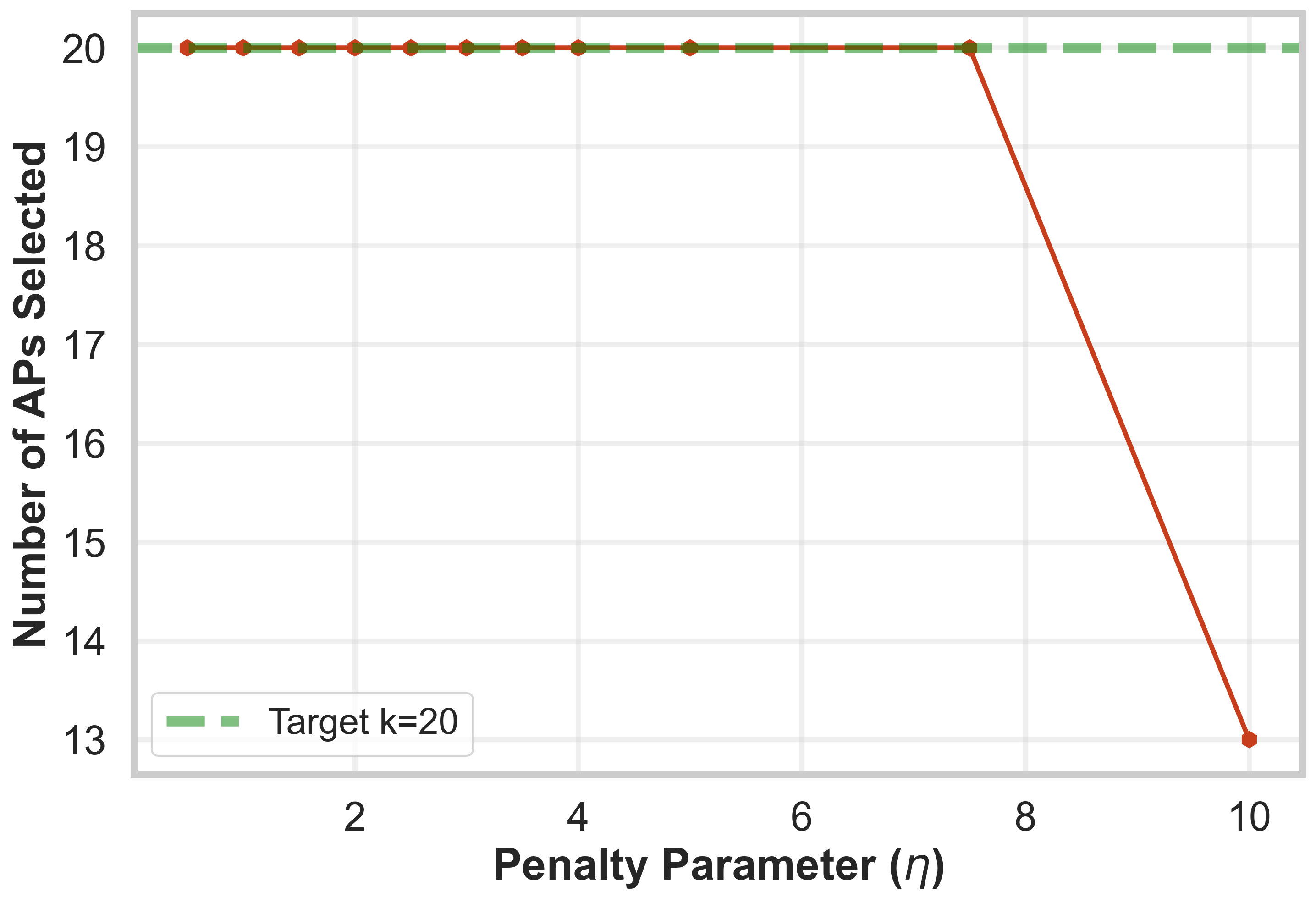}
    \caption{\texorpdfstring{$\eta$}{eta} Impact on the Number of Selected APs.}
    \label{fig:num_aps_selected_Penalty}
\end{figure}

\begin{figure}[!t]
    \centering
    \includegraphics[width=0.99\linewidth]{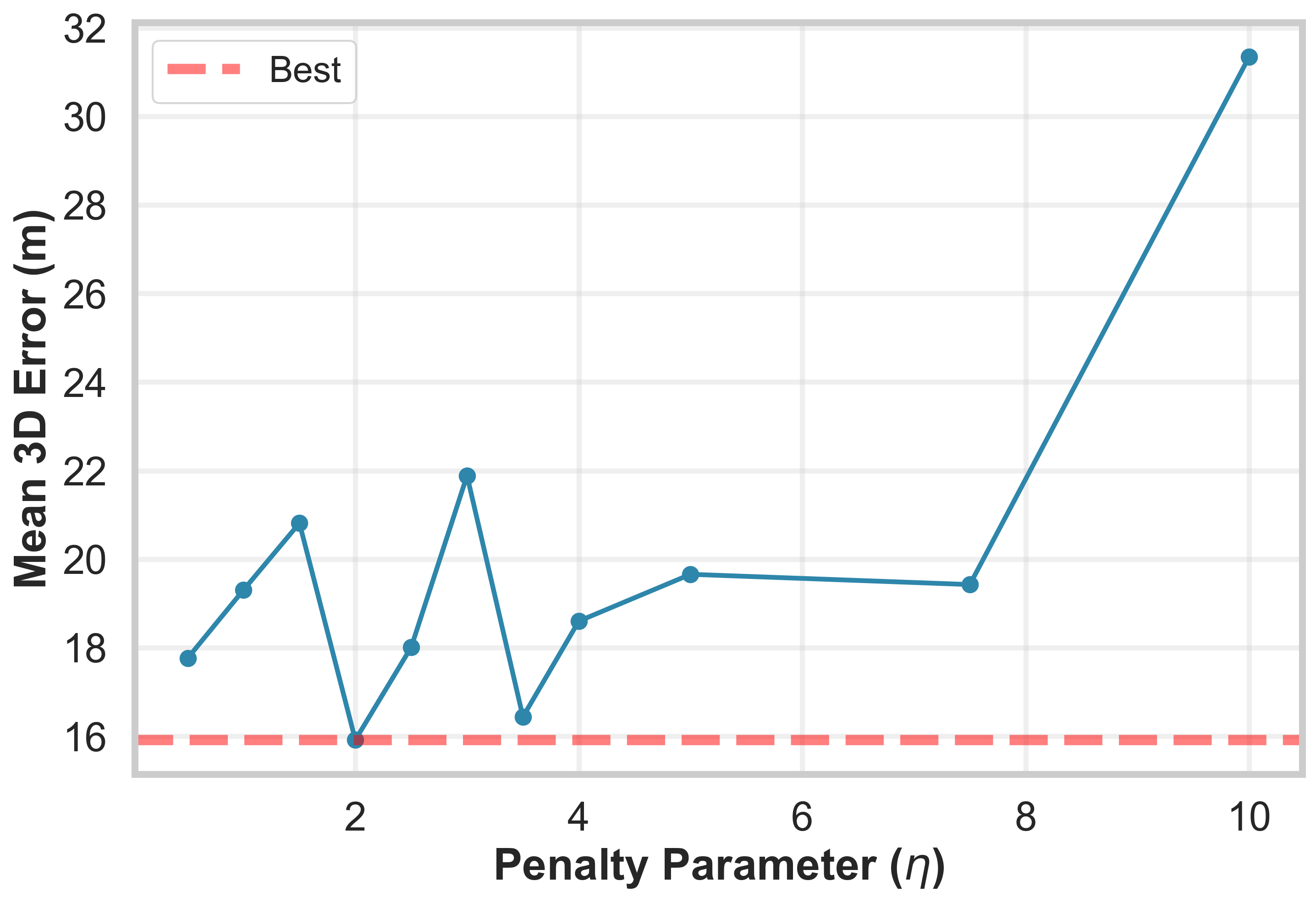}
    \caption{\texorpdfstring{$\eta$}{eta} Impact on the 3D Mean Error.}
    \label{fig:3D_mean_error_Penalty}
\end{figure}

Figure~\ref{fig:num_aps_selected_Penalty} shows the effect of increasing the penalty parameter $\eta$ on the selected number of APs (at budget $k=20$). Figure~\ref{fig:3D_mean_error_Penalty} further shows the 3D localization accuracy. The figures show that, for small values of the penalty, the optimizer can reach the target of $k$ APs, with a minimum localization accuracy at $\eta=2$. For larger penalty values, the constraint term creates an overly harsh energy landscape, causing the annealer to get trapped in local minima before reaching the target number of APs $k$, which in turn affects the localization accuracy.

\subsubsection{Effect of The Budget Parameter \texorpdfstring{$k$}{k}}

\begin{figure}[!t]
    \centering
    \includegraphics[width=0.99\linewidth]{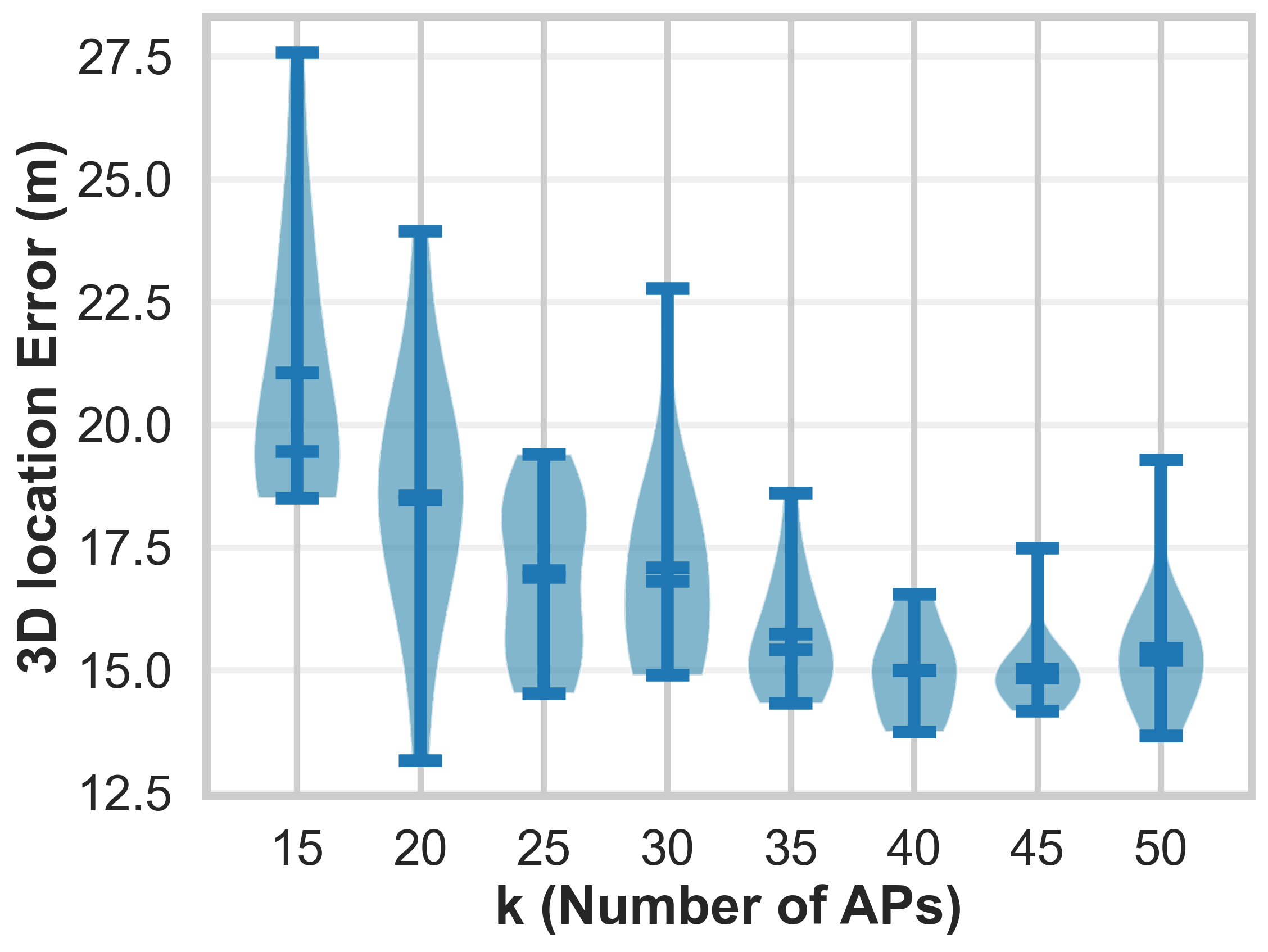}
    \caption{Effect of increasing budget $k$ on the 3D localization error.}
    \label{fig:error_vs_k_violin}
\end{figure}

Figure~\ref{fig:error_vs_k_violin} illustrates the relationship between the AP budget $k$ and the resulting localization error. As anticipated, increasing the budget (i.e., selecting more APs) consistently improves localization accuracy. This improvement, however, comes at the cost of increased computational complexity in the positioning process.

\subsection{Effect of Annealing Parameters}

\subsubsection{Effect of Inverse Temperature Parameter \texorpdfstring{$\beta$}{beta}}
\begin{figure}[!t]
    \centering
    \includegraphics[width=0.99\linewidth]{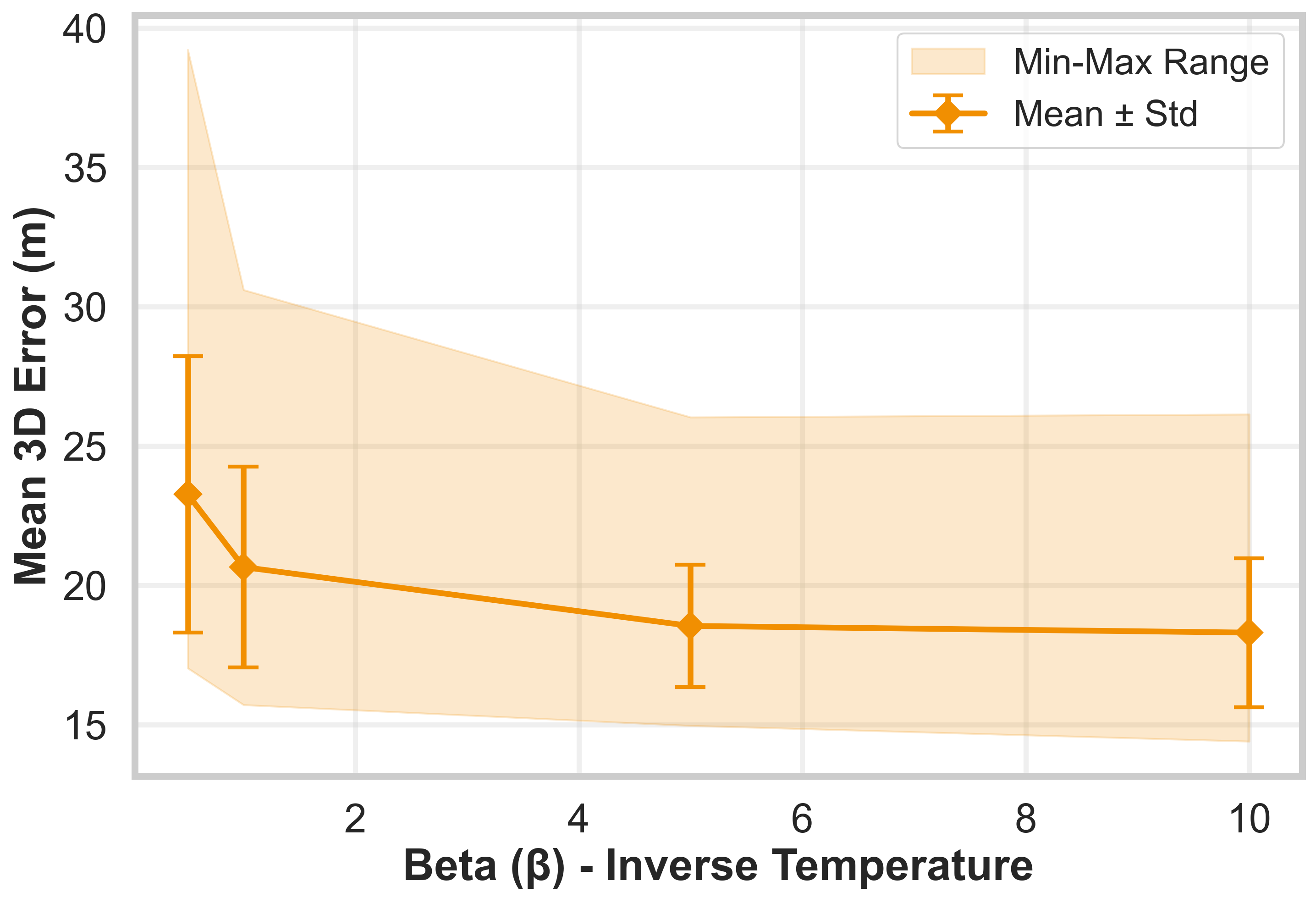}
    \caption{Effect of $\beta$ on mean 3D error.}
    \label{fig:beta_vs_error}
\end{figure}

Figure~\ref{fig:beta_vs_error} demonstrates the impact of the inverse temperature parameter $\beta$ on 3D localization accuracy in the quantum annealing process. In this context, $\beta$ functions as a scaling factor that controls the influence of the problem's energy landscape, effectively determining how sharply the solver distinguishes between optimal and suboptimal solutions. The figure shows that localization accuracy improves as $\beta$ increases because a higher value helps the annealer to more effectively navigate the complex cost function and settle into a deeper, more optimal energy state corresponding to a better AP selection. However, this improvement saturates beyond a value of $\beta = 5$.

\subsubsection{Number of Sweeps}
\begin{figure}[!t]
    \centering
    \includegraphics[width=0.99\linewidth]{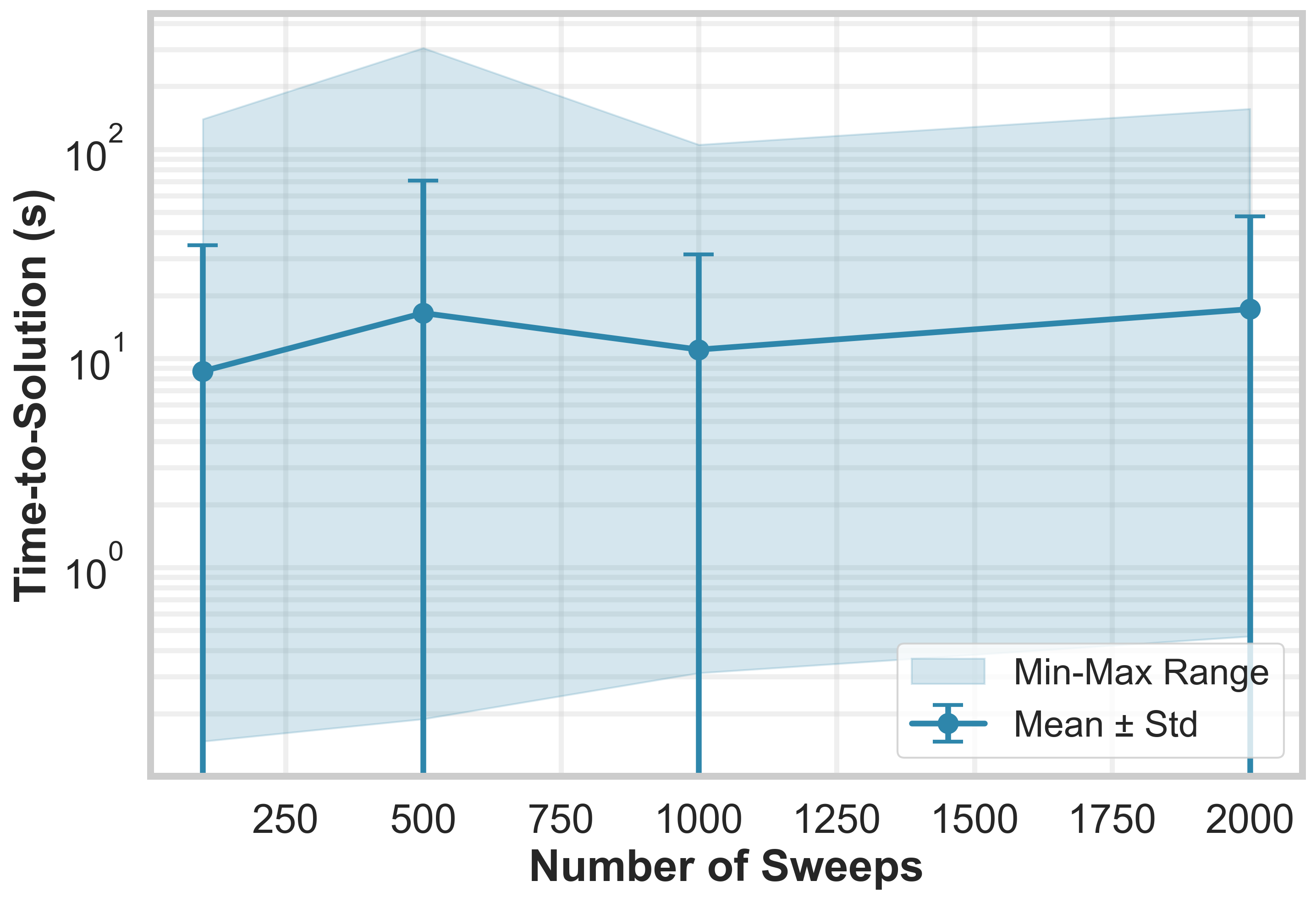}
    \caption{Effect of sweeps on time-to-solution (TTS).}
    \label{fig:sweeps_vs_tts}
\end{figure}

Figure~\ref{fig:sweeps_vs_tts} illustrates the effect of the number of sweeps on the time-to-solution (TTS). TTS represents the total time required to successfully solve the problem with a high probability, encompassing annealing time, the number of iterations, and all associated overhead. As expected, the results show that a higher number of sweeps directly leads to an increased TTS. This relationship is intuitive because each sweep represents a fundamental update step in the annealing schedule. Allocating more sweeps allows the solver to more thoroughly explore the solution space, which naturally requires more computational time to complete.

\subsection{Comparison with Classical APs Selection}

\subsubsection{3D Localization Accuracy}

\begin{figure}[!t]
    \centering
    \includegraphics[width=0.95\linewidth]{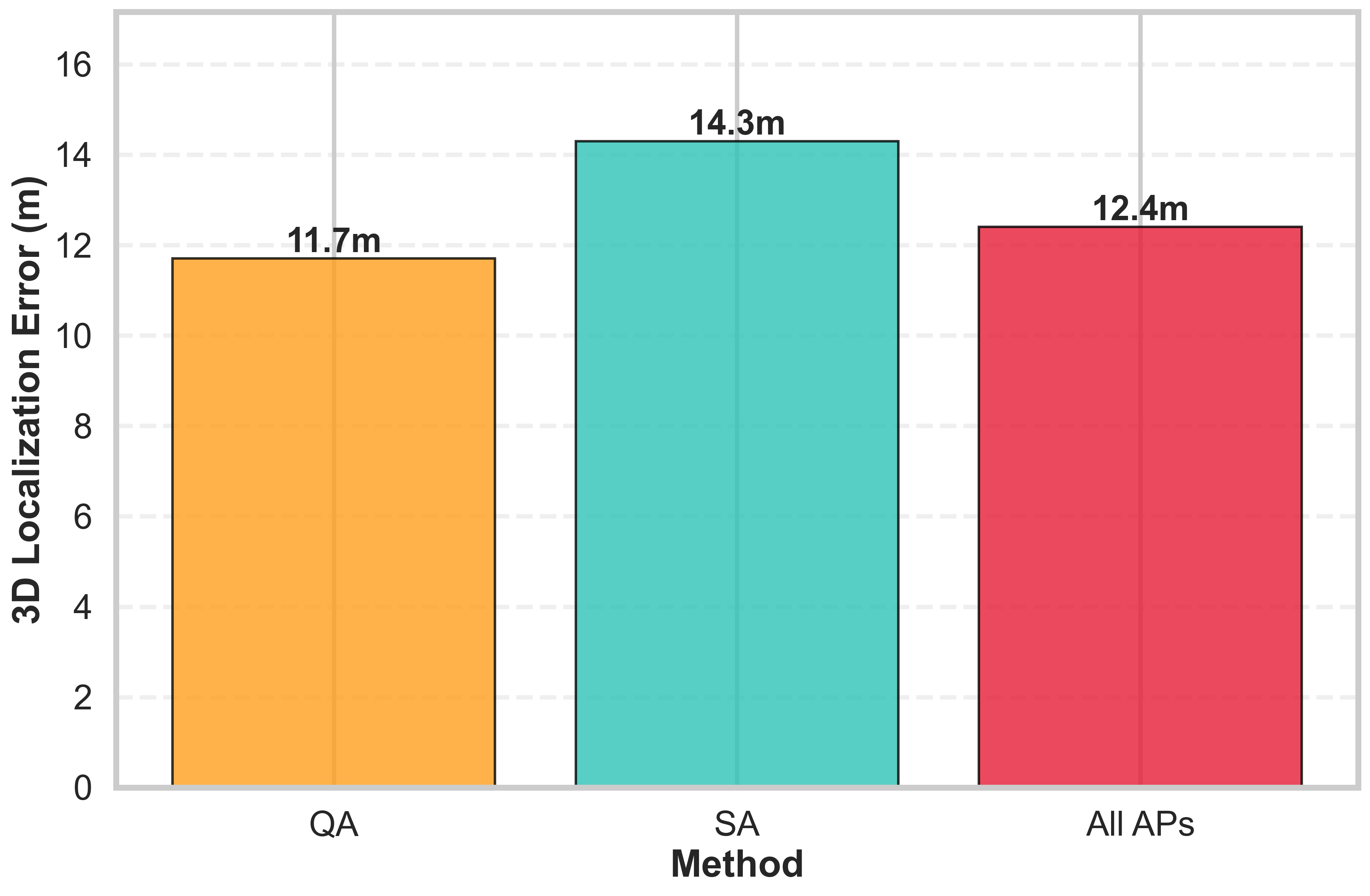}
    \caption{3D Localization Error Between Different Methods.}
    \label{fig:boxplot_3d_error}
\end{figure}

Figure~\ref{fig:boxplot_3d_error} presents the mean 3D localization error across the different AP selection methods. The figure shows that the quantum annealing approach (QA) achieves a mean error of \textbf{11.7m}, which is lower than simulated annealing (SA) at \textbf{14.3m} and lower than using all APs having error of \textbf{12.4m}.
This demonstrates that our quantum AP selection selects a more effective subset than classical optimization methods, identifying truly informative APs while eliminating redundancy.

\subsubsection{Floor Accuracy}
\begin{figure}[!t]
    \centering
    \includegraphics[width=0.99\linewidth]{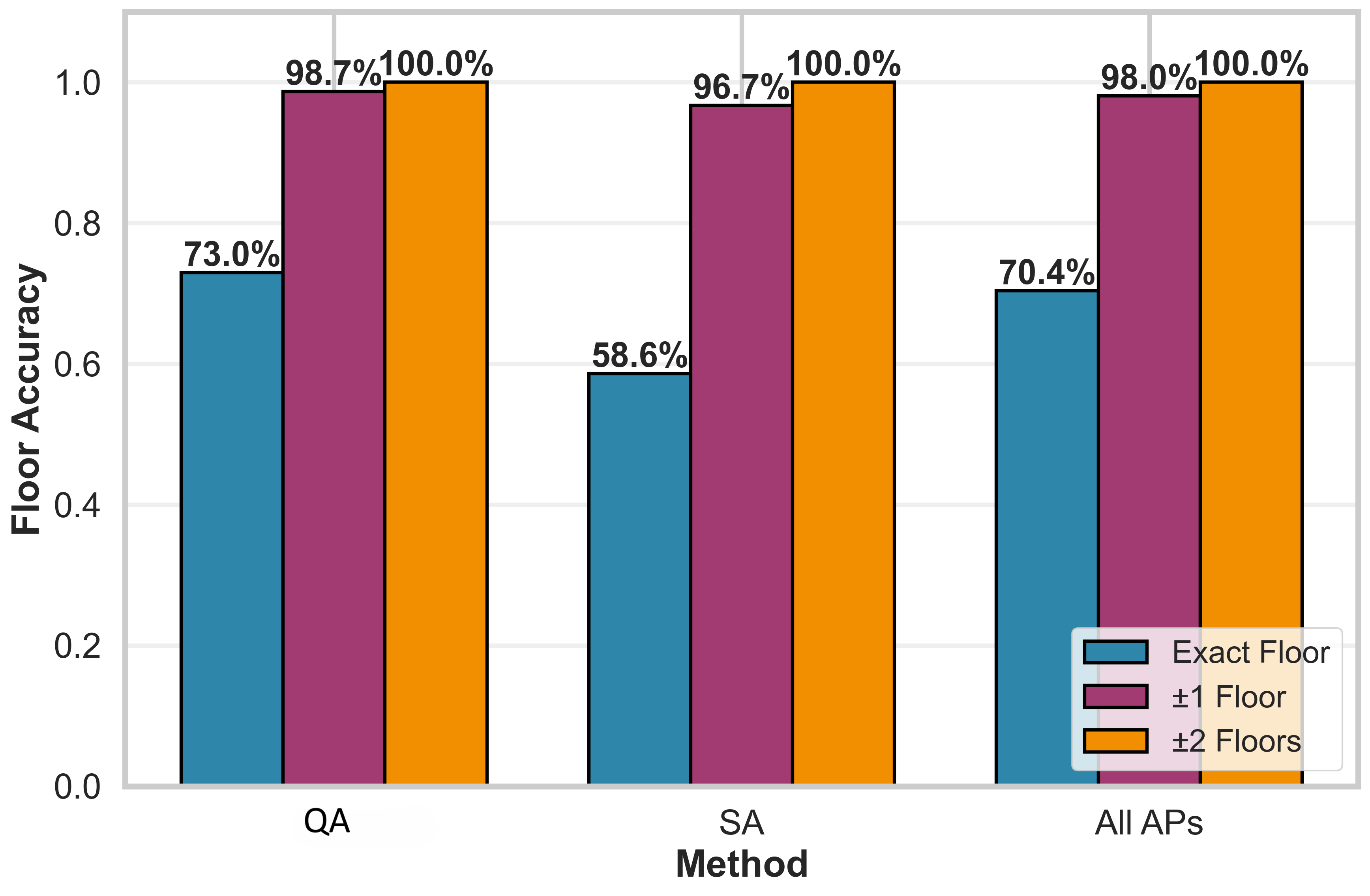}
    \caption{Floor Localization Accuracy Comparison.}
    \label{fig:floor_localization_sa_qa_all_aps}
\end{figure}

Figure~\ref{fig:floor_localization_sa_qa_all_aps} shows a comparison between the different APs selection techniques for floor localization. The figure shows that the floor localization accuracy of the quantum annealing approach (\textbf{73\%}) is better than the simulated annealing (\textbf{58.6\%}). Moreover, it is better than the accuracy obtained by using all the APs (\textbf{70.4\%}). This is because quantum annealing eliminates the redundant and noisy APs in the environment.

\subsubsection{Annealing Time}

\begin{figure}[!t]
    \centering
    \includegraphics[width=.95\linewidth]{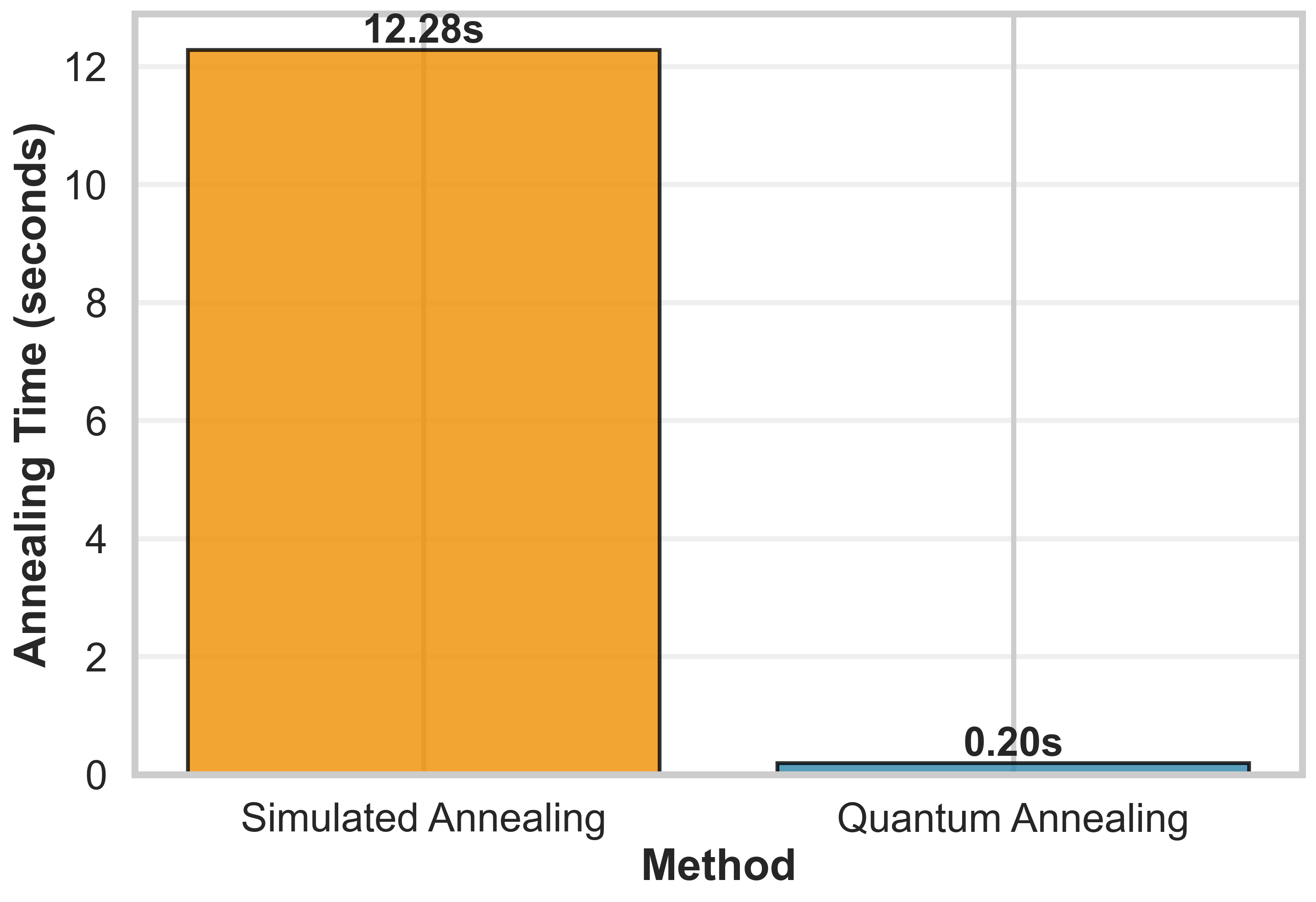}
    \caption{Annealing Time Comparison.}
    \label{fig:annealing_time}
\end{figure}

Figure~\ref{fig:annealing_time} compares the annealing time of the classical simulated annealing against the proposed quantum annealing. The figure shows that quantum annealing achieves a significantly lower annealing time, demonstrating its efficiency in solving our QUBO problem compared to classical simulated annealing.

\section{Conclusion}
\label{sec:conclude}
This paper presented a novel quantum algorithm for selecting Access Points (APs) under a predefined budget constraint for indoor localization. Our approach was based on formulating the AP selection challenge as a Quadratic Unconstrained Binary Optimization (QUBO) problem, which was then solved using quantum annealing to determine the optimal AP subset without exceeding the budget. This methodology offered a pathway to significantly lower infrastructure costs with only a minimal effect on localization performance.

The algorithm was implemented and evaluated within a comprehensive 3D testbed. The experimental findings demonstrated a dramatic reduction in infrastructure needs, achieving a \textbf{96.1\%} decrease in the number of APs required while maintaining comparable 3D localization accuracy. The proposed quantum method also surpassed classical selection techniques, delivering superior accuracy and a substantial computational speedup. It achieved a mean time-to-solution of \textbf{0.20 seconds}, which is $61\times$ faster than the classical benchmark, all while reducing the mean localization error by \textbf{10\%} compared to the classical simulated annealing. For floor localization, the quantum approach achieved \textbf{73\%} accuracy, outperforming both the classical AP selection (\textbf{58.6\%}) and even the complete set of APs (\textbf{70.4\%}). These results highlight the significant potential of our quantum AP selection framework for practical, large-scale 3D localization.

\bibliographystyle{IEEEtran}
\bibliography{references}

\end{document}